\documentclass[
reprint,
amsmath,
amssymb,
aps, physrev,
floatfix,
]{revtex4-2}

\usepackage{times}
\usepackage{amsthm}
\usepackage{eucal}
\usepackage{xcolor}
\usepackage[normalem]{ulem}
\usepackage{mathrsfs}
\usepackage{amsmath}
\usepackage{graphicx}% Include figure files
\usepackage{dcolumn}% Align table columns on decimal point
\usepackage{bm}% bold math

\newcounter{movie}

\begin{document}
% \preprint{APS/123-QED}
%\title{Statistical Correlations of Topological Defects in a Model Monolayer Confluent Tissue}

\title{\textbf{Topological Defects Mediate Collective Transport of Confluent Cells}}
\author{Jiusi Zhang$^{1}$}
\author{Chung Wing Chan$^{1}$}
\author{Bo Li$^{2,3}$}
\author{Rui Zhang$^{1,4,5}$}
\email{ruizhang@ust.hk}
\affiliation{$^{1}$Department of Physics, Hong Kong University of Science and Technology, Clear Water Bay, Kowloon, Hong Kong SAR\\$^{2}$Institute of Biomechanics and Medical Engineering, Applied Mechanics Laboratory, Department of Engineering Mechanics, Tsinghua University, Beijing 100084, PR China\\$^{3}$Mechano-X Institute, Department of Engineering Mechanics, Tsinghua University, Beijing 100084, PR China\\$^{4}$State Key Lab of Displays and Opto-electronics, Hong Kong University of Science and Technology, Clear Water Bay, Kowloon, Hong Kong SAR\\$^{5}$Center for AI for Science, Hong Kong University of Science and Technology, Clear Water Bay, Kowloon, Hong Kong SAR}%

\date{\today}% It is always \today, today,
             %  but any date may be explicitly specified

\begin{abstract}
Collective cell migration governs a range of physiological and pathological processes, from tissue morphogenesis to cancer invasion, in which topological defects arise as an inevitable consequence of frequent cellular rearrangement and migration. Here, we employ an Active Vertex Model to investigate structural defects generated in the wake of transported cells. We find that while the drag coefficient of a cell in a perfect lattice is anisotropic, the threshold drag force required to mobilize the cell is isotropic. Remarkably, we find that dragging two neighboring cells along the direction of least-resistance minimizes lattice disruption. By comparing defect-healing behaviors across different physical models, we disentangle the contributions of cell adhesion and many-body interactions. Together, our findings provide new insights into the topological organization of confluent tissues during collective migration, advancing our physical understanding of cellular transport processes such as wound healing, tissue repair, and cancer metastasis.
%There is a recent interest in elucidating topological defects in confluent cells, yet the role of defects during collective cell transport remains elusive. Here, we employ an Active Vertex Model to investigate structural defects generated in the wake of transported cells. We find that while the drag coefficient of a cell in a perfect lattice is anisotropic, the threshold drag force required to mobilize the cell is isotropic. Remarkably, we find that dragging two neighboring cells along the direction of least-resistance minimizes lattice disruption. By comparing defect-healing behaviors across different physical models, we disentangle the contributions of cell adhesion and many-body interactions. Together, our findings provide new insights into the topological organization of confluent tissues during collective migration, advancing our physical understanding of cellular transport in processes such as wound healing and cancer metastasis.
\end{abstract}

\maketitle

\section{\label{sec:level1}Introduction}

Confluent cells represent a unique type of soft material, which comprises closely packed, active, and deformable biological cells~\cite{RN126, RN130}. Over the past decade, there has been a growing multidisciplinary interest in studying these confluent cell systems~\cite{RN57, RN58, RN59}. From a physicist's perspective, confluent cells can be considered as a form of active matter~\cite{RN138}, where the collective dynamics of self-propelled cells can lead to the emergence of various spatiotemporal patterns~\cite{RN134, RN135, RN19, RN136, RN141, RN150}. This perspective has provided valuable insights into many biological processes, including embryonic development~\cite{RN44}, wound healing~\cite{RN41, RN44, RN163, RN248}, cancer progression~\cite{RN44, RN40, RN55, RN101}, and cell fate dynamics~\cite{RN10}. 
A better physical understanding of structural dynamics and rheological properties of confluent cells may also enhance our abilities to engineer tissues for applications such as tissue regeneration, transplantation, and scar removal~\cite{RN149}.

A characteristic physical property of confluent cells is their motility-mediated liquid--solid transition~\cite{RN127, RN129, RN133}. The melting transition is also termed epithelial--mesenchymal transition~\cite{RN204, RN162}, and the reverse transition is alternatively called glass transition~\cite{RN3, RN33, RN127}. T1 transitions---topological rearrangements of neighboring cells---can enable tissue remodeling and fluidization by allowing tissue to flow while preserving cell--cell junctions~\cite{RN204}. Theory and experiment have demonstrated that the liquid--solid transition is influenced by cellular activity, intracellular mechanics, surface tension, and cell--cell interactions~\cite{RN3, RN6, RN45, RN7, RN26, RN24, RN47, RN255}. For example, confluent cells typically exhibit fluid-like behavior when intercellular adhesion outweighs cellular contraction, particularly when combined with elevated cell motility~\cite{RN190, RN3, RN6}.

Monolayer confluent cells can be represented by polygons, and this foam-like structure underlies their collective dynamics. Cells with more than and fewer than 6 neighbors are structurally irregular cells, and can be called positive ($+$) and negative ($-$) disclinations (defects) in the lattice structure, respectively. These defect cells are not merely structural irregularities destroying hexagonal lattice order but may hold biological significance, potentially playing an important role in embryo development, tissue functioning, and morphogenesis~\cite{RN123, RN125, RN151, RN183, RN184, RN198}, and their dynamics have been shown to govern the ordering kinetics and collective migration of endothelial cell layers~\cite{RN234}. Between the solid and liquid phases, the tissue monolayer can enter a hexatic phase, in which quasi-long-range orientational order governs both its mechanical and structural properties~\cite{RN137, RN139, RN112, RN111}. 

The anisotropy and deformability of confluent cells are also key determinants of their structure and dynamical behavior. A monolayer of confluent cells can be viewed as a two-dimensional (2D) active nematic liquid crystal, in which neighboring anisotropic cells tend to align with one another~\cite{RN42, RN43, RN19, RN135, RN239, RN248}. $+1/2$ and $-1/2$ disclinations (defects)---the hallmarks of such 2D nematics---have been identified in the coarse-grained cell deformation field~\cite{RN182} from a diverse range of biological systems~\cite{RN117, RN191, RN4, RN135, RN192}. The collective motion of cells generates mechanical stresses and spontaneous flows around these singular defects~\cite{RN51, RN11}, and also mobilizes $+1/2$ defects. Consequently, such defects appear to influence collective dynamics during tissue morphogenesis~\cite{RN38, RN35, RN43, RN248} and to regulate cell fate, including apoptosis and extrusion~\cite{RN10, RN23}.

The direction of self-propulsion of $+1/2$ defects in these living systems can serve as an indicator of whether the active stress is extensile (pusher-like) or contractile (puller-like). However, many tissues composed of contractile cells exhibit extensile behavior, as $+1/2$ defects tend to propel toward their head. This puzzling phenomenon has been attributed to polar fluctuating forces~\cite{RN4, RN51, RN21, RN11} and to the role of adhesion proteins~\cite{RN118}.

More general topological defects, such as p-atic defects, have been analyzed in confluent cells~\cite{RN25}. For example, it was found in a simulation that cell--cell alignment effects can lead to concentrations of pentagonal cells at the cores of $+1$ nematic defects~\cite{RN110}. In another simulation work, the unbinding of hexatic ($\pm1/6$) defects are \textcolor{blue}{is} found to be responsible for T1 transitions of confluent cells, in which structural defects can accumulate in the wake of migrating cells and progressively destroy the hexatic order of the tissue~\cite{Krommydas_2025}. Recent works have focused on elucidating correlations between differently defined defects in steady-state confluent cells~\cite{RN115}. Nevertheless, the roles various types of defects play in collective cellular transport phenomena, including morphogenesis, wound healing, and cancer invasion, are still poorly understood~\cite{RN173, RN174, RN175}.

To address this open question, here we use the Active Vertex Model (AVM) to analyze the behaviors of topological defects in cellular transport phenomena. We focus on structural defects and nematic defects in this work. We first examine their statistics in liquid and solid phases. We then perform cell-drag simulation to model active microrheology measurement. Specifically, by simulating a cell dragged through a hexagonal lattice, we show that cells must overcome an isotropic energy barrier to migrate despite the lattice anisotropy. We further analyze defect pattern in the wake of dragged cell's trajectory. Remarkably, we find that dragging two neighboring cells can substantially minimize defects in their wake. Topological defects can self-heal outside a packet of migrating cells, which exhibit a nontrivial periodic walking-like dynamic pattern. Similar defect-healing phenomena are also observed in fundamentally different models, including colloidal crystals with Yukawa potentials, with Lennard--Jones potentials, and the NexTissUe model~\cite{RN163}, a sophisticated particle-based model of interacting cells. By comparing self-healing behaviors across these distinct models, we uncover the effects of cell adhesion, repulsion, and many-body interactions of cells on defect healing processes. 
%Our results offer new physical insights into defect-free collective migration of cells in biological tissues.

%\textcolor{blue}{Most importantly, we demonstrate that dragging two neighboring cells can autonomously heal structural defects in their wake — an self-healing phenomenon driven by a neutral defect packet executing a nontrivial periodic walking pattern. We find that this mechanism is closely associated with the intercellular repulsion, the core structure of the defect packet, the short-ranged attraction, and the T1 transition energy barrier. The robustness of this phenomenon across fundamentally different model systems — the AVM, Yukawa and Lennard-Jones colloidal systems, and the NexTissUe model — demonstrates that self-healing is a universal feature of hexatic solid mechanics, offering a new physical perspective on defect-free collective migration in biological tissues.}

\section{\label{AVM}Active Vertex Model}

% figure 1 
\begin{figure}[t]
\includegraphics[width=\linewidth]{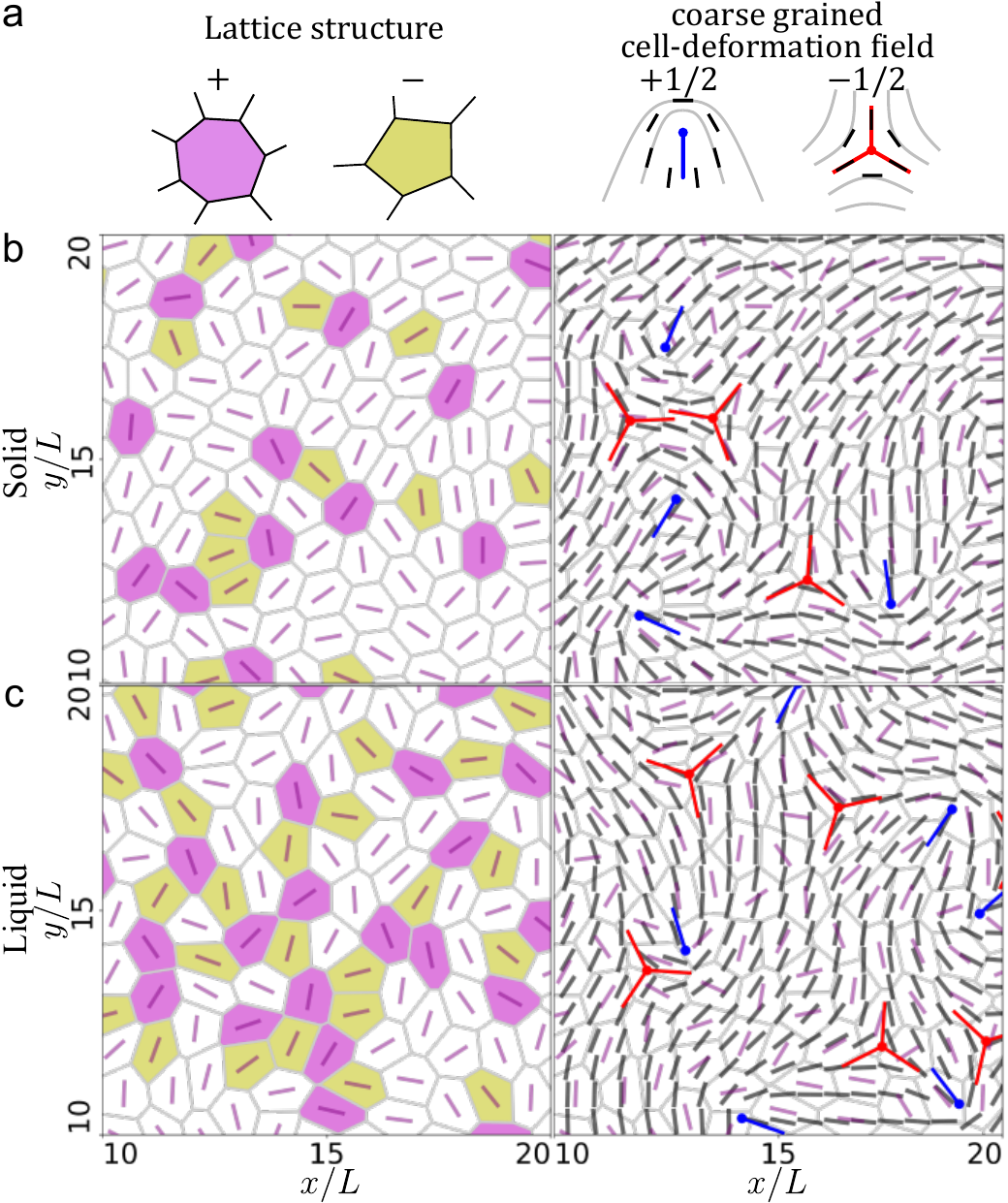}% Here is how to import EPS art
\caption{\label{fig:1} \textbf{Topological defects in the AVM model.} \textbf{a} Representative pictures of structural defects and nematic defects. \textbf{b} A snapshot of the system (left) and the corresponding director field (right) in the solid state with $\bar{v}_0=0.3$ and $p_0=3.55$. \textbf{c} A snapshot of the system (left) and the corresponding director field (right) in the liquid state with $\bar{v}_0=0.3$ and $p_0=3.85$. $+$ and $-$ defect cells are marked purple and yellow, respectively, and $+1/2$ and $-1/2$ defects are marked by blue lines and red trefoils, respectively.
}
\end{figure}

To investigate the physical interactions among confluent cells, various 2D computational models have been developed to effectively describe phenomena such as glass transition, cell migration, and cell mechanics~\cite{RN58, RN59}. These models include Self-Propelled Voronoi (SPV) model~\cite{RN3, RN52}, Active Vertex model (AVM)~\cite{RN6, RN246}, multiphase field model~\cite{RN32, RN26,RN115}, NexTissUe model~\cite{RN163}, Active Tension Network model~\cite{RN48}, mesoscopic model~\cite{RN51}, and cellular Potts model~\cite{RN112, RN33, RN112}. Here we focus on AVM to investigate the structure of confluent tissues subjected to active and driven motions of cells~\cite{RN246, RN150}, as it offers a minimum model with detailed representations of the cellular structures and well-defined physical boundaries. Additionally, its computational efficiency makes it suitable for large-scale simulations, which are achieved by retaining the essential interactions necessary to capture the many-body interactions among cells.

We construct a 2D network of cell vertices to represent a confluent cellular system consisting of $N=900$ cells undergoing active motions in a $30\times30$ box with periodic boundary conditions in both directions (Supplementary Movie 1). According to the previous works~\cite{RN185, RN190, RN186, RN7, RN6}, the vertex motion is governed by the AVM free energy consisting of a surface area elasticity term, a contraction term, and a cell--cell adhesion energy term (SI):
\begin{equation}\label{eq:AVM}
    E_\mathrm{AVM} = \sum_{i = 1}^{N}\left[\frac{K}{2}(A_i - A_0)^2 + \frac{\Gamma}{2}P_i^2\right] + \Lambda\sum_{(\mu, \nu)} l_{\mu\nu},
\end{equation}
where $K$, $\Gamma$, and $\Lambda$ are the elasticity, contraction, and adhesion moduli, respectively~\cite{RN190}, $A_0$ is the target cell area, $A_i$ and $P_i$ are the area and perimeter of cell $i$, and $l_{\mu\nu}$ is the length of the edge connecting vertices $\mu$ and $\nu$. The position of vertex $\mu$ evolves according to ~\cite{RN186}
\begin{equation}\label{eq:motion}
    \frac{\mathrm{d} \mathbf{r}_\mu}{\mathrm{d} t} = \frac{1}{\zeta} \mathbf{F}_\mu + v_0 \sum_{i\in C_\mu}\hat{\mathbf{n}}_i,
\end{equation}
where $\zeta$ is the friction coefficient, $\mathbf{F}_\mu$ is the force on vertex $\mu$ derived from $E_\mathrm{AVM}$, $C_\mu$ is the set of cells sharing vertex $\mu$, and $\hat{\mathbf{n}}_i = (\cos\theta_i, \sin\theta_i)$ is the self-propulsion direction of cell $i$, where the angle $\theta_i$ follows a Wiener process with a rotational diffusivity $D_r$ (SI).

The behavior of the system is tuned by two key parameters: (i) to incorporate cell motility in the AVM~\cite{RN6}, each cell is self-propelled at a constant speed $v_0$ along a freely rotating direction characterized by angle $\theta_i$; (ii) parameter $p_0$ sets the target shape of the cells via~\cite{RN190}
\begin{equation}\label{eq:p0}
    p_0 = -\frac{\Lambda}{\Gamma\sqrt{A_0}},
\end{equation}
which is a direct consequence of the competition between cell adhesion and contraction (SI).

We express all quantities in terms of fundamental dimensions. The length unit $L$ is chosen such that the target area of a cell is $A_0=1$, which gives rise to a lattice constant of the perfect hexagonal lattice in the simulation to be $a = \sqrt{\frac{2 A_0}{\sqrt{3}}} \approx 1.075 L$. For conciseness, the unit $L$ may be omitted in the following discussions. By choosing an arbitrary time unit $\tau_0$, the velocity unit is given by $L/\tau_0$. The dimensionless activity parameter $\bar{v}_0\equiv v_0\tau_0/L$ will be adopted in what follows. In our simulation, $\bar{v}_0$ is varied within $0<\bar{v}_0<1$, while the shape parameter $p_0$ is constrained to $3.3 < p_0 < 4$.

We then identify topological defects from the simulation data and perform calculations over the statistics of these defects (Fig.~\ref{fig:1}). Structural defects are easy to identify. To extract nematic defects, we determine the elongation direction using the shape tensor $\mathbf{f}_i$ and identify the positions and orientations of the defects; numerical details are provided in SI~\cite{RN11, RN119, RN239, RN89}. This physics-based computational approach yields results that are consistent with our recently developed AI-based defect detection method~\cite{RN119}.

\begin{figure*}
\centering
\includegraphics[width=\linewidth]{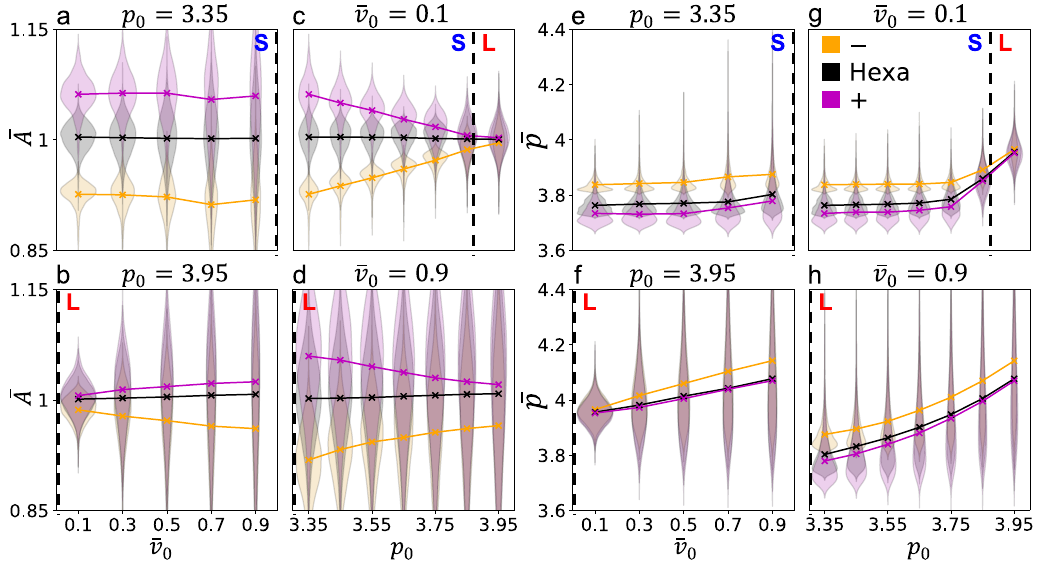}
\caption{\label{fig:L} \textbf{Statistics of different types of cells.} (\textbf{a}--\textbf{d}) Violin plots of the average cell area $\bar{A}$ for $+$ defect, $-$ defect, and hexagonal cells. (\textbf{e}--\textbf{h}) Violin plots of the average shape-parameter $\bar{p}$ for $+$ defect, $-$ defect, and hexagonal cells. The dashed lines mark the transition point between the solid and liquid phase.
}
\end{figure*}

\section{Results}

\subsection{\label{subsec:level1}Structural defect analysis}

The number of neighboring cells for a cell, namely cell neighbor (coordination) number $z$, is an important quantity to characterize the structure of the tissue~\cite{RN125, RN50}.
%, as the local spatial correlations between cells can significantly influence the structure and rheology of a tissue and the morphogenesis of an embryo. 
For a 2D confluent tissue, the ground state configuration is a honeycomb lattice with $z\equiv 6$ for every cell. Cells with greater or fewer neighbors are called structural defects or disclinations. Here, we measure the ensemble average of the cell area $\bar{A}=\langle \frac{1}{N}\sum_{i=1}^N{A_i}\rangle$ and the shape parameter $\bar{p}=\langle \frac{1}{N}\sum_{i=1}^N{p_i}\rangle$ in steady state as functions of different parameters for the two types of defect cells and the hexagonal cells in Fig.~\ref{fig:L}.
We find that compared to the reference hexagonal cells with $z=6$, $+$ defect cells with $z>6$ are larger and rounder, while $-$ defect cells with $z<6$ are smaller and more elongated (Figs.~\ref{fig:L},~S1%\ref{fig:L_3d}
).
In other words, cells with fewer neighbors will favor a more anisotropic shape. This can be understood by the fact that the statistical variation of the intercellular forces, which dictates the shape of the cell of interest, depends on the number of edges---a smaller number of edges implies a larger variation, which can lead to a more anisotropic cell shape. 
The fact that $+$ defect cells are larger in area $\bar{A}$ is because their shape, with more edges, is closer to a circle, which maximizes the area for a given perimeter.

Although $\bar{v}_0$ and $p_0$ play a similar role in dictating the phase of the confluent cells as demonstrated in the phase diagram of the AVM model~\cite{RN3, RN111}, their impacts on the different types of cells are different.
By varying the activity parameter $\bar{v}_0$ with a fixed shape parameter $p_0$, we find that the difference in the average area $\bar{A}$ between the three types of cells is insensitive to $\bar{v}_0$ in the solid state (Fig.~\ref{fig:L}a), but increases as $\bar{v}_0$ increases in the liquid state (Fig.~\ref{fig:L}b). This implies that the solid state is insensitive to the magnitude of the cell activity, as cells are trapped in their local positions and their activity acts similar to thermal noise. In the liquid state, however, cells with different $z$ exhibit different mobilities, as cells with fewer (more) neighbors can squeeze more (fewer) at higher activity, leading to smaller (larger) $\bar{A}$ (Fig.~\ref{fig:L}b).
If we fix $\bar{v}_0$ but vary $p_0$, the trend becomes independent of the phase of the system: the difference in $\bar{A}$ among the three types of cells diminishes as $p_0$ increases (Fig.~\ref{fig:L} c and d). A higher $p_0$ can promote deformations and rearrangement of all types of cells, which can homogenize them, giving rise to less differences in their $\bar{A}$'s.

\begin{figure*}[t]
\includegraphics[width=\linewidth]{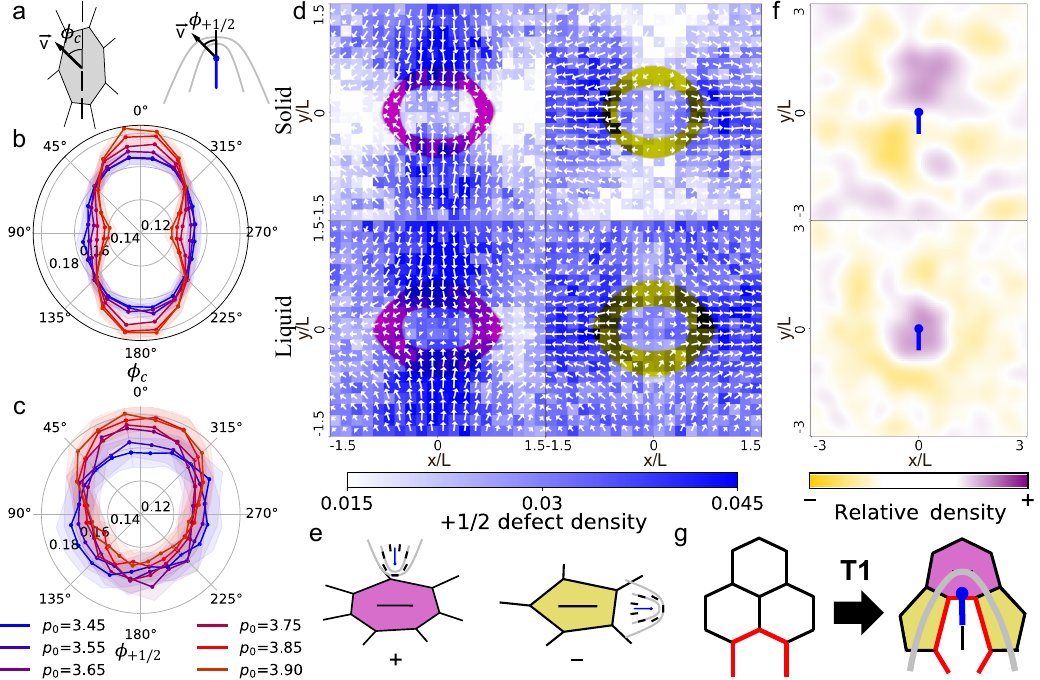}
\caption{\label{corrlationfig}\textbf{ Statistics of cells and topological defects.} \textbf{a} Schematic definitions of two angles. Left: angle between a cell's velocity vector and its long axis, $\phi_{c}$; right: angle between a $+1/2$ defect's velocity vector and its orientation direction, $\phi_{+1/2}$. \textbf{b} Rose-plot of the histogram of $\phi_{c}$ at $\bar{v}_0 = 0.3$. \textbf{c} Rose-plot of the histogram of $\phi_{+1/2}$ at $\bar{v}_0 = 0.3$. Different colored curves are for different $p_0$. The liquid--solid transition point is $p_0 \approx 3.7$. \textbf{d} Heatmap of the distribution of $+1/2$ defects around an average $+$ defect cell (purple) and an average $-$ defect cell (yellow) in the solid (upper) and liquid (lower) state. The white arrows indicate the average orientation of the $+1/2$ defects at that point. \textbf{e} Schematic diagrams showing how $+1/2$ defects tend to orient and spatially distribute around $\pm$ defects. \textbf{f} The distribution of $\pm$ defects around an average $+1/2$ defect in the solid ($\bar{v}_0=0.3$, $p_0=3.55$) and liquid ($\bar{v}_0=0.3$, $p_0=3.85$) state. \textbf{g} A schematic explanation of the spatio-orientational correlation between $\pm$ defects and $+1/2$ defects by looking at how a cell invades an otherwise hexagonal lattice.
}
\end{figure*}

The behavior of the average cell-shape parameter $\bar{p}$ is similar to that of the average cell area $\bar{A}$ (Fig.~\ref{fig:L} e--h). But there is a general trend in $\bar{p}$: as the system moves towards a more liquid-like state, $\bar{p}$ among all three types of cells will increase, because cells become more elongated and are also more homogeneous in the more liquid-like state. Because the total area of all cells is a conserved quantity, there is no general trend for the change in the $\bar{A}$ plot (Fig.~\ref{fig:L} a--d).

Our analyses also reveal that the hexatic phase emerges at the onset of the solid--liquid transition, characterized by a peak in the hexatic order and a corresponding minimum in the structural defect density (Fig.~S2%\ref{fig:N_OL&hexaOrder}
)~\cite{RN139}. Below the transition, limited cell motility limits the generation and unbinding of defects and preserves the hexagonal order, while above the transition, increased activity disrupts the spatial order, leading to a rise in the defect density (Figs.~S2%\ref{fig:N_OL&hexaOrder}
,~S3%\ref{fig:L_count}
,~S4%\ref{fig:N_6&Order}
). This behavior is consistent with previous theoretical and numerical studies on the hexatic phase~\cite{RN139, RN111, RN112, RN114, RN137}. Interestingly, the number density of $\pm1/2$ defects does not exhibit any non-monotonic, critical-like behavior around the transition point. This indicates that the nematic order derived from the coarse-grained deformation field is not sensitive to the hexatic order of confluent cells, but depends continuously on the parameters $v_0$ and $p_0$.

Analysis of the statistics of the lifetimes of structural defects reveals that their distribution follows an exponential function. We find that the half-life of structural defects, namely $t_{1/2}$, decreases toward zero as the system moves toward the liquid state, in which defect proliferation and annihilation events occur more frequently (Figs.~S5%\ref{fig:lifetime}
~and~S6%\ref{fig:HF}
).

\subsection{\label{subsec:level2}Self-propulsion of cells and $+1/2$ defects}
To understand the collective dynamics of confluent cells, we next focus on the statistics of self-propelled cells and $+1/2$ defects. We introduce two angles, $\phi_c$ and $\phi_{+1/2}$, to represent the velocity angle of a cell and a $+1/2$ defect, respectively (Fig.~\ref{corrlationfig}a). As shown in Fig.~\ref{corrlationfig}b, cells prefer to move along their elongation axis, despite the inherent randomness in their self-propulsion directions. Since cell velocity is determined by a combination of its activity and cell--cell interaction forces, this directional bias suggests that cells experience reduced resistance from neighboring cells along their elongation axis compared to the transverse direction. 
This effect becomes particularly pronounced deep in the liquid state, in which cells are, on average, more elongated.
In certain cell monolayers, such as Madin-Darby Canine Kidney (MDCK) epithelial cells, there is a robust correlation between the direction of intercellular stress and the instantaneous cell velocity---a phenomenon known as plithotaxis~\cite{RN195, RN196}. This anisotropic stress field also dictates cellular morphology, inducing local orientational order in cell elongation. Thus, intercellular stress provides the mechanistic link connecting the preferred direction of cell velocity to the axis of cell elongation. More importantly, this establishes that certain cell monolayers can be regarded as liquid crystals~\cite{RN195, RN196}.

A similar analysis on $+1/2$ defects shows that they tend to propel along their head direction, and this trend is pronounced in the liquid state with large $p_0$ (Fig.~\ref{corrlationfig}c). This implies that confluent cells in the AVM model can be mapped to an extensile active LC---this result is consistent with a theoretical prediction~\cite{RN11}. But our work points out that this extensile character is weakened in the solid state with small $p_0$. This $p_0$ dependence of the defect behavior echoes another recent experimental finding that cell--cell adhesion is responsible for the extensile behavior of the tissue~\cite{RN4}. We further show that $+1/2$ defects in the liquid state exhibit enhanced extensile character when they age (Fig.~S7%\ref{fig:lifetime_plus+1/2}
). This is because newly generated defects have to overcome the attractive force by the oppositely-charged counterpart defects, and thereby exhibit weak extensile behavior. As they age, they unbind and their self-propulsion becomes more directional.

\begin{figure*}[t]
\includegraphics[width=\linewidth]{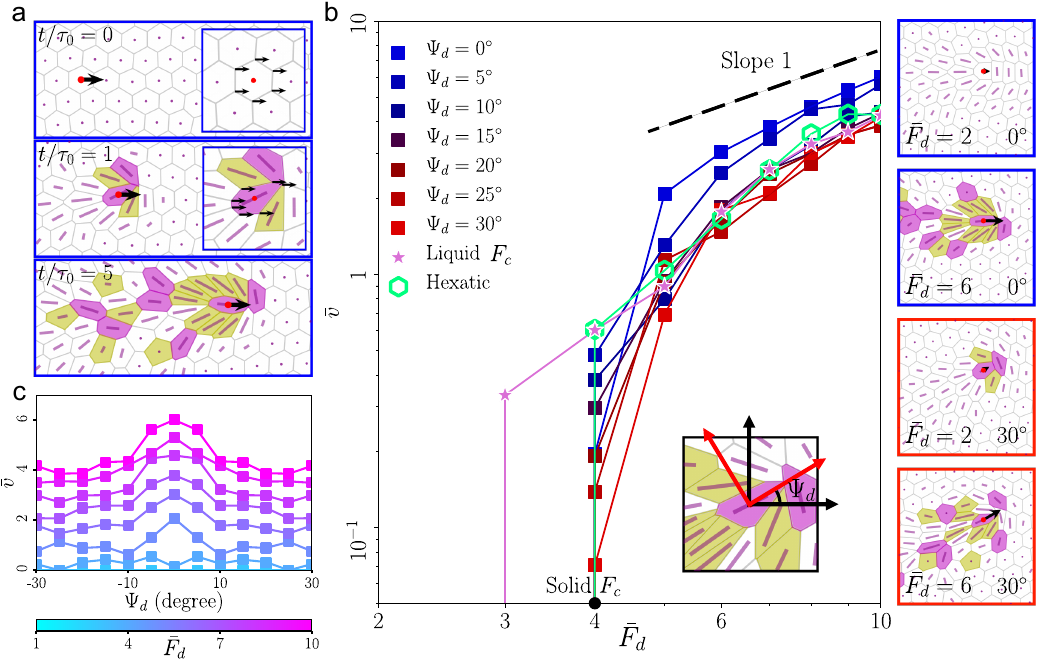}% Here is how to import EPS art
\caption{\label{fig:4-1} \textbf{Single-cell drag simulation.} \textbf{a} Snapshots of the system when a cell is dragged with $\bar{F}_\text{d}=6$ and $\Psi_\text{d}=0^\circ$. Insets: cell dragging is realized by applying additional forces to its vertices. \textbf{b} The velocity--force relation for the dragged cell at different force angles in the solid state ($\bar{v}_0=0.1$ and $p_0=3.65$), the hexatic state ($\bar{v}_0=0.3$ and $p_0=3.65$), and liquid state ($\bar{v}_0=0.5$ and $p_0=3.65$); right four panels: steady-state snapshots of the system for different parameter sets $(\bar{F}_\text{d},\Psi_\text{d})$. The threshold force for both the solid and hexatic state is $\bar{F}_\text{c}\simeq 4$. \textbf{c} The average velocity $\bar{v}$ of the dragged cell plotted against force angle $\Psi_\text{d}$ for different $\bar{F}_\text{d}$.
}
\end{figure*}

\subsection{\label{subsec:level3}Correlation between structural and nematic Defects}

Next, we investigate the statistical correlation between structural defects and nematic defects. By analyzing the probability distribution of $+1/2$ defects around an average $+$ or $-$ defect cell in either solid or liquid state, we observe different correlation patterns (Fig.~\ref{corrlationfig}d). In the solid state, $+1/2$ defects are predominantly located on the two sides of a $+$ defect cell, oriented towards the cell, whereas near a $-$ defect cell they are more likely to appear in the two end regions, facing outwards (Fig.~\ref{corrlationfig}e). In the liquid state, this correlation pattern is weakened due to frequent exchanges of neighboring cells.

The correlation between the two types of defects can alternatively be characterized by the probability distribution of $\pm$ defect cells around an average $+1/2$ defect (Fig.~\ref{corrlationfig}f). In the solid state, $+$ defect cells are predominantly located in the head region of the $+1/2$ defect (purple), while $-$ defect cells are more likely to be found in the tail region (yellow) (Fig.~\ref{corrlationfig}f). In the liquid state, however, $+$ defect cells are more frequently observed in the core regions of the $+1/2$ defect. 

These spatio-orientational correlation patterns can be qualitatively understood by an illustrative picture shown in Fig.~\ref{corrlationfig}g. When a cell invades an otherwise hexagonal lattice, the cell in the immediate front of the invading cell gains an additional neighbor and also becomes elongated in the orthogonal direction due to the mechanical force exerted by the invader. Meanwhile, cells on the two sides of the invading cell, initially in contact, are now pushed apart, losing coordination number $z$ and at the same time being stretched along the invasion direction. Therefore, this process induces the formation of a $+1/2$ defect, with its head area occupied by $+$ defect cells and tail area filled with $-$ defect cells (Fig.~\ref{corrlationfig}e). This picture well rationalizes the observation that $+1/2$ defects tend to stay on either side of $+$ defects and face towards them, but prefer to stay near the end region of $-$ defects and face backwards.

The above statistical findings are further confirmed by several confluent cell experiments~\cite{RN10, RN25, RN4} (Fig.~S8%\ref{fig:exp evi}
). 
%These results are also consistent with the findings of Rozman et al., who reported a spatial correlation between $+1$ defects and $-$L cells with $z=5$~\cite{RN110}. 
A recent study using a multiphase field model also produced similar statistical correlations between structural and nematic defects~\cite{RN115}, further confirming that this spatio-orientational correlation is model independent.

\begin{figure*}[t]
\includegraphics[width=\linewidth]{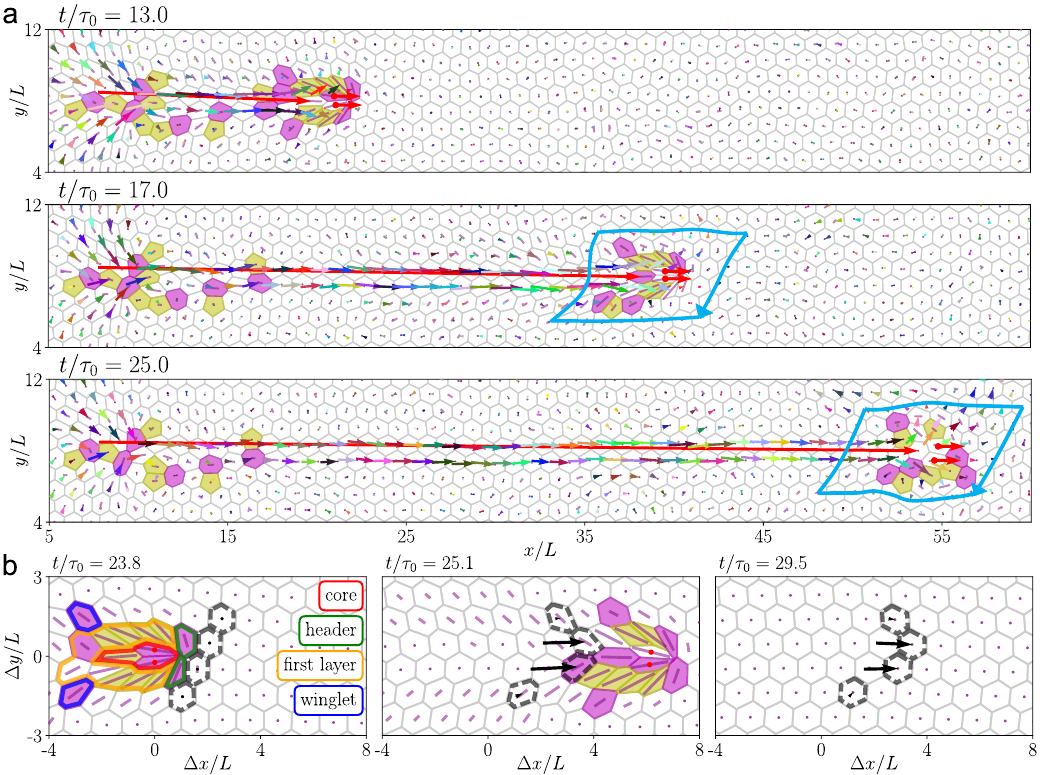}
\caption{\label{fig:dragconf} \textbf{Double-cell drag simulation.} \textbf{a} Snapshots of dragging two neighboring cells at $\Psi_\text{d}=0^\circ$, $\bar{F}_\text{d}=6$. The blue vectors are the Burger's vectors around a packet of the moving cells containing structural defects. To clearly distinguish the displacement vector field of the cells, we use random colored arrows pointing from the initial position at $t/\tau_0=10$ to the current position. \textbf{b} Snapshots of the packet undergoing self-healing of structural defects, along with the displacement vectors for black-outlined cells across different layers before and after the packet passes through.}
\end{figure*}

\begin{figure*}[t]
\centering
\includegraphics[width=.7\linewidth]{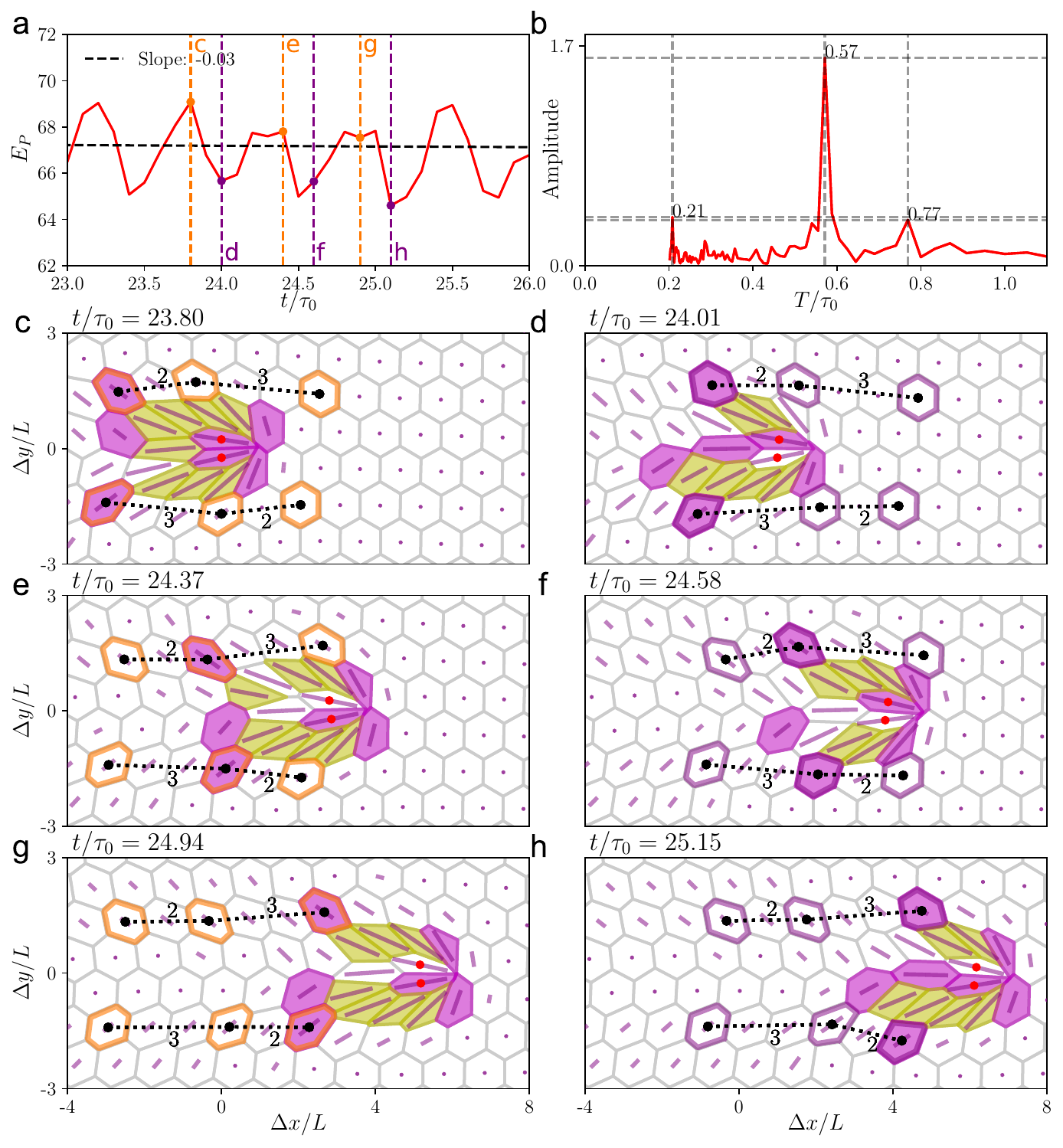}
\caption{\label{fig:peakvalley_snapshot} \textbf{Periodicity observed in the double-cell drag simulation.} \textbf{a} Temporal behavior and \textbf{b} the Fourier analysis of the contraction and adhesion energy $E_\mathrm{P}$ at $\bar{v}_0 = 0.01$ and $p_0 = 3.65$. Snapshots of the packet for (\textbf{c}, \textbf{e} and \textbf{g}) peak points and (\textbf{d}, \textbf{f} and \textbf{h}) valley points in the $E_\mathrm{P}$ plot. The positions of the winglet cells are highlighted with orange and purple outlines for the peak and the valley state, respectively.}
\end{figure*}

\subsection{\label{subsec:level4}Dragging one cell}

To investigate how cell migration impacts confluent tissues and to probe their rheology, we conduct cell-drag simulations. As shown in Fig.~\ref{fig:4-1}a, a cell is dragged by a constant external force $F_\text{d}$ in a hexagonal lattice in the solid state at low activity ($\bar{v}_0 = 0.1$, $p_0 = 3.65$). The dimensionless force $\bar{F}_\text{d}\equiv F_\text{d} \tau_0/(\zeta L)$ is used in the subsequent discussion. To account for the six-fold symmetry of the lattice, the force is applied at various angles, $\Psi_\text{d}$, with respect to the $x$-axis, ranging from $0^\circ$ to $30^\circ$. 

When $\bar{F}_\text{d}$ is below a certain threshold $\bar{F}_\text{c}\equiv F_\text{c}\tau_0/(\zeta L)$, the cell remains trapped near its initial position due to the blocking of the surrounding cells. Notably, $\bar{F}_\text{c}$ is independent of the force angle $\Psi_\text{d}$, indicating an isotropic free energy barrier despite the anisotropy of the hexagonal lattice. When $\bar{F}_\text{d} \ge \bar{F}_\text{c}$, the cell moves persistently along a specific direction (Fig.~\ref{fig:4-1}, Fig.~S9%\ref{fig:DL}
). The liquid state exhibits a reduced threshold $\bar{F}_\text{c} \simeq 3$, which grows to $\bar{F}_\text{c} \simeq 4$ as the system solidifies into the hexatic and solid phases, reflecting the deeper caging energy as neighbor exchange becomes increasingly constrained--- this is consistent with the picture that caging strength grows progressively as the system solidifies toward the glass transition~\cite{RN236}.

We further observe that the mean velocity $\bar{v}$ of the dragged cell subjected to a fixed $\bar{F}_\text{d}$ is a function of the force angle $\Psi_\text{d}$ (Fig.~\ref{fig:4-1} b and c). The cell moves fastest when $\Psi_\text{d} = 0^\circ$, at which it tends to travel along a ``zigzag'' tunnel between two horizontal layers of cells (Supplementary Movie 2). In contrast, the cell moves slowest at $\Psi_\text{d} = 30^\circ$. At this unfavorable force angle, the dragged cell cannot find a viable pathway without colliding with other cells. This drag-coefficient anisotropy vanishes in the liquid state (Supplementary Movie 3).

During the continuous dragging, the cell experiences resistance due to cell--cell interactions (excluded-volume repulsion and adhesion). By tracking the trajectory of the dragged cell in all directions, we observe that the migration angle may deviate from the force angle $\Psi_\text{d}$, exhibiting a weak directional locking effect~\cite{DLscience} (Fig.~S9%\ref{fig:DL}
). This occurs because cells experience the least drag along $0^\circ$, causing them to migrate towards this ``locking'' direction.

When $\Psi_\text{d} = 30^\circ$, the dragged cell experiences the highest resistance. Due to the symmetry of the lattice, it still migrates at $30^\circ$ on average.
Along the orthogonal direction, however, we observe that the cell exhibits super-diffusive behavior, that is, the mean square displacement along that direction scales as $\sim t^{\alpha}$, where $\alpha \simeq 1.15$ (Fig.~S10%\ref{fig:MSD}
). 

By monitoring the velocity of the dragged cell, we further observe that it exhibits persistent stick-slip motion when $\bar{F}_\text{d} > \bar{F}_\text{c}$, which can be characterized by intermittent trapping within cages (evidenced by speed dropping to zero, cf Fig.~S11%\ref{fig:stick-slipmotion}
). This stick-slip behavior persists over a wide range of conditions within the regular hexagonal lattice, contrasting with homogeneous, amorphous tissues where the stick-slip behavior can transition into a more continuous motion when $\bar{F}_\text{d}$ is well above $\bar{F}_\text{c}$~\cite{RN26}. 
The stick-slip motion mode breaks down at $\Psi_\text{d}=0^\circ$ in our simulation, at which the dragged cell moves with fluctuating but nonzero velocity, exhibiting continuous locomotion without the ``stick'' mode (Fig.~S11%\ref{fig:stick-slipmotion}
). This can be understood by the fact that at this special force angle, the dragged cell can easily sneak through two horizontal layers of cells.

We also observe that the dragged cell tends to elongate into a $+$ defect (i.e., gain neighbors), with its elongation direction parallel to its migration direction (Fig.~\ref{fig:4-1}a). Cells in its immediate front tend to gain neighbors, become stretched in the orthogonal direction, and form $+$ defects. Therefore, a $+1/2$ defect can be seen in front of the dragged cell (Fig.~\ref{fig:4-1}a). This simultaneous formation of $+1/2$ and structural defects can also be understood by the spatio-temporal correlation found in the steady state (Fig.~\ref{corrlationfig}). Interestingly, the illustrative picture of a cell invading a hexagonal lattice can also be applied to here to understand the formation of defects around the dragged cell (Fig.~\ref{corrlationfig}g).

\subsection{\label{subsec:level5}Dragging two cells}

\begin{figure}
\includegraphics[width=\linewidth]{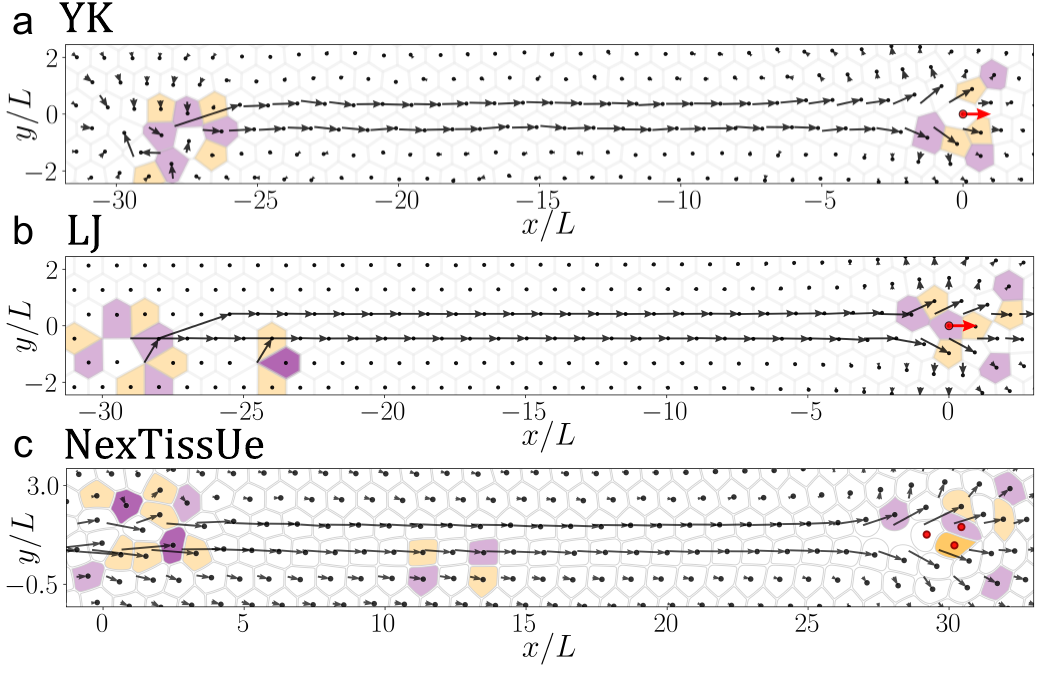}% Here is how to import EPS art
\caption{\label{fig:double_drag_models}\textbf{Double-row self-healing configurations in different models.} Double-row self-healing in \textbf{a} a colloidal monolayer with Yukawa (YK) interactions at $q_\text{d}/q_0 = 5$, \textbf{b} in a colloidal monolayer with Lennard--Jones (LJ) interactions at $\epsilon_\text{d0}/\epsilon_\text{00}\approx7.746$, and \textbf{c} in a cell monolayer using NexTissUe model. In all three systems, the majority of structural defects heal back into hexagons, producing two adjacent rows of cells with a $1a$ displacement. The particles or cells displaced over $5a$ are marked by red dots without showing the displacement vector.
}
\end{figure}

\begin{figure*}[t]
\includegraphics[width=\linewidth]{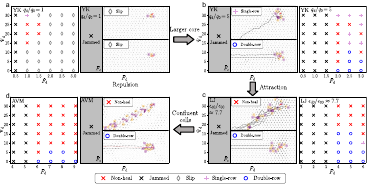}
\caption{\label{fig:Mechanism_PD} 
\textbf{Defect-healing mode diagram and representative snapshots in different models.} \textbf{a} The slip mode dominates in a wide range of parameters for the repulsive YK system; \textbf{b} A larger dragging core ($q_\text{d}/q_0=5$) in the YK model favors multi-row self-healing and disrupts the slip mode, with double-row self-healing appearing near $0^\circ$ and single-row self-healing favored near $30^\circ$; \textbf{c} The LJ system containing a short-range attraction with $\epsilon_\text{d0}/\epsilon_\text{00}\approx7.746$ disrupts single-row self-healing and favors multi-row defect healing at $0^\circ$; \textbf{d} Compared to the LJ model, the AVM suppresses complete defect healing for large $\Psi_\mathrm{d}$ due to cell confluence and many-body interaction of cells.
}
\end{figure*}

During the dragging of a single cell, structural defects are generated in the wake of its trajectory, and they do not spontaneously annihilate and restore the hexatic order; instead, they form a persistent trace of $\pm$ defects that disrupt the hexagonal lattice structure (Figs.~\ref{fig:4-1},~S12%\ref{fig:dragconfigs}
a). To look for the possibility of healing these wake defects, we have explored various cell-dragging scenarios, including dragging an additional cell over the wake region, back-and-forth dragging of a cell, and dragging two cells in various ways (Fig.~S12%\ref{fig:dragconfigs}
, Supplementary Movie 4).
Among these, remarkably, we find that defect-healing can happen when two neighboring cells are dragged together along $\Psi_\text{d}\simeq0^\circ$ (Supplementary Movie 5). As shown in Fig.~\ref{fig:dragconf}, defect cells are only present around the initial and final locations of the two cells.

We next analyze the lattice structure during the driven motion of two neighboring cells. As the two cells advance, we observe a group of $\pm$ defects surrounding them, and they form a compact cluster that exhibits a highly regular pattern, which we term them ``self-healing packet'' (Fig.~\ref{fig:dragconf}a). As this packet advances, its Burgers vectors amount to zero, indicating topological charge neutrality of structural defects within this region (Fig.~\ref{fig:dragconf}a). However, this packet contains five more cells than a perfect hexagonal lattice occupying the same region (Fig.~\ref{fig:dragconf}a). Among these additional cells, two of them are the dragged cells, and one cell right behind them, namely the ``follower cell'', is co-moving with them and exhibits a large displacement vector (Fig.~\ref{fig:dragconf}a). These three cells form the core of the packet and persistently migrate together within the tissue (Fig.~\ref{fig:dragconf}b). The remaining two additional cells are dynamic: during the motion of the packet, two rows of cells advance collectively, each displacing by one lattice constant $a$ forward per cycle, while all other cells return to their initial locations (Fig.~\ref{fig:dragconf}a and b). The displacement vector field has a net flux into the packet, and contributes to the two additional cells.

%flux at the boundary of the packet is therefore proportional to the total number of additional cells, and can be decomposed into two contributions: long-range displacement vectors from the three fixed core cells, and short-range displacement vectors of magnitude $1a$ from the two dynamic cells.

Within the packet, two cells in front of the dragged cells tend to form $+$ defects, which we term them header cells. On the two sides of the three core cells, $-$ defects appear and we call them the first-layer cells. These cells can also be seen in the single-cell drag simulation (Fig.~\ref{fig:dragconf}b). To compensate for the topological charge of the packet, $+$ defects tend to form in the tail of the first-layer cells, with another two $+$ defect cells located in the second (outer) layer, which we term winglet cells (Fig.~\ref{fig:dragconf}b, Fig.~S13%\ref{fig:two_dynamics}
). As will be elaborated below, these charge-compensating defect cells change dynamically during the motion of the packet and mediate in the self-healing of $pm$ defects in the wake region (Fig.~S13%\ref{fig:two_dynamics}
).

We further analyze the temporal behavior of the system and find that the spectrum of the contraction and adhesion energy, defined as $E_\mathrm{P} = \sum^{N}_{i=1} \frac{\Gamma}{2} (P_i - P_0)^2$, exhibits a pronounced peak at $T_{f_1}=2.5T_0$, where $T_0$ is the characteristic time scale for a dragged cell to advance by one lattice constant $a$ (Fig.~\ref{fig:peakvalley_snapshot} a and b, Figs.~S14%\ref{fig:energyterms}
,~S15%\ref{fig:fourier}
). By examining the snapshots of the packet at the peak and valley points of the $E_\mathrm{P}$ plot, we observe an interesting periodic-like behavior of the packet structure, in which the winglet cells play an important role. As the system evolves from a peak (valley) point to the next peak (valley) point, the two winglet cells located on the sides of the packet will advance by $\sim 2a$ (Fig.~\ref{fig:peakvalley_snapshot} c and e) and $\sim 3a$ (Fig.~\ref{fig:peakvalley_snapshot} d and f), respectively; during the next period, however, the two cells will advance in an alternate manner, i.e., by $\sim 3a$ (Fig.~\ref{fig:peakvalley_snapshot} e and g) and $\sim 2a$ (Fig.~\ref{fig:peakvalley_snapshot} f and h), respectively. Therefore, after two periods, the packet represented by the winglet cells will advance by $5a$. This gives rise to the emerging period $T_{f_1}=2.5T_0$ for the advancing packet. During the motion of the packet, $\pm$ defect cells in the wake can self-heal in a periodic manner while the winglet cells are advancing (SI)%(Appendix~\ref{sec:mechanism})
. The nontrivial period of $2.5T_0$ can also be found in the temporal behavior of other quantities (Fig.~S15%\ref{fig:fourier}
), such as the coordination number $z$ of the follower cell (Fig.~S15%\ref{fig:fourier}
h). 

Also note that this self-healing behavior is robust against simulation parameters. Under variations of the background activity $\bar{v}_0$, force magnitude $\bar{F}_\text{d}$, and force angle $\Psi_\text{d}$, we can still observe a perfect lattice in the wake of the trajectories of the two dragged cells (Fig.~S16%\ref{fig:phaseDiagram}
). 
The phenomenon is also insensitive to the choice of the integration time step $\delta t$ and the threshold length $l_\text{min}$ for T1 transition, as long as the simulation time resolution is sufficiently high to faithfully resolve cell rearrangements during T1 transitions (Fig.~S17).

%\ref{fig:phaseDiagram_delta_t_lmin})

%The robustness further extends to the numerical integration parameters: the self-healing configuration is preserved as long as the cell displacement per timestep remains below the T1 transition threshold length $l_\text{min}$, ensuring that neighbor exchanges are resolved faithfully rather than being artificially triggered by excessively large time steps. As shown in Fig.~S17%\ref{fig:phaseDiagram_delta_t_lmin}
%, the self-healing configuration is unaffected across a wide range of $\delta t$ and $l_\text{min}$ values satisfying this condition, confirming that the observed phenomenon is a robust physical feature of the confluent tissue.

\subsection{\label{differentsystems}Self-healing phenomena in other systems}
Here we ask: why defects can self heal and is the self-healing of defects observed in the AVM model unique to the specific model and to confluent cells? To address this question and to further elucidate the mechanism underlying the defect healing phenomenon, we have performed particle or cell drag simulations using other models. We choose two model systems, namely the colloidal particle model and the NexTissUe model~\cite{RN163}, with details described below. 

A monolayer of colloidal particles can be mapped onto a confluent cell system by processing Voronoi cell partition. Therefore, the colloidal particles provide the simplest model on which we can elucidate the effects of cell--cell interactions on defect-healing. For this model, we consider two different inter-particle potentials: the Yukawa (YK) potential between particle $i$ and $j$ in the reduced units reads~\cite{RN237}
\begin{equation}
V_{\text{YK}}(r_{ij}) = \frac{q_i q_j}{r_{ij}} e^{-\kappa r_{ij}},
\end{equation}
where $q_i$ and $q_j$ are the charges of particle $i$ and $j$, respectively, $\kappa^{-1}$ sets the length scale of the screened Coulomb interaction, and $r_{ij}$ is the separation distance between the two particles.
The Lennard--Jones (LJ) potential between particle $i$ and $j$ reads~\cite{RN259}
\begin{equation}
V_{\text{LJ}}(r_{ij}) = 4\epsilon_{ij}\left[\left(\frac{\sigma_{ij}}{r_{ij}}\right)^{12} - \left(\frac{\sigma_{ij}}{r_{ij}}\right)^6\right],
\end{equation}
where $\sigma_{ij}$ and $\epsilon_{ij}$ are the location and the depth of the potential well quantifying the strength of their attraction.
The cross-species parameters follow the Lorentz--Berthelot mixing rule $\sigma_{ij} = (\sigma_{ii}+\sigma_{jj})/2$ and $\epsilon_{ij} = \sqrt{\epsilon_{ii}\,\epsilon_{jj}}$~\cite{Berthelot, RN263}. The LJ potential contains both short-range repulsion and longer-range attraction forces, whereas the YK model only has repulsive interactions. 
In the YK model, we choose charge $q_0=1$ for all particles except the dragged particle carrying charge $q_\text{d}$. Similarly in the LJ model, we choose $\epsilon_{00}=0.05$ for all particles except the dragged particle having a different energy scale $\epsilon_\text{d0}/\epsilon_\text{00}$.
The NexTissUe model is a sophisticated cell model, in which each cell is represented by a deformable ring-polymer that self-propels via a polarized lamellipodium and reorients its polarity through contact inhibition of locomotion, and unlike the strictly confluent AVM, the model permits finite gaps between cells (SI)~\cite{RN252}.

In all three different models, we have observed various types of defect-healing phenomena (Fig.~\ref{fig:double_drag_models}). In particular, by properly choosing system parameters in these models, we can reproduce the ``double-row'' healing of defects found in the AVM model, that is, two rows of cells will advance collectively with the dragged cells, and almost all lattice defects between the initial and current locations of the dragged cells will be eliminated (Fig.~\ref{fig:double_drag_models}). The very existence of the auto-healing phenomenon in these distinct models underscores the robustness and generality of our discovery of defect-free transport of cells.

%In all three systems -- colloidal particles with Yukawa (YK) interactions~\cite{RN237}, colloidal particles with Lennard-Jones (LJ) interactions %(Appendix~\ref{sec:colloidal_sim})
%, and the NexTissUe model~\cite{RN163} (SI) -- the majority of $\pm$ defects heal back into hexagons, producing two adjacent displaced cell rows with a $1a$ displacement, suggesting a universal underlying mechanism.  Despite this fundamental difference in model construction, double-row self-healing is clearly observed (Fig.~\ref{fig:double_drag_models}c), providing strong evidence for the robustness of the phenomenon in confluent tissue systems.

We further analyze the dynamic details of defect-healing in the cell/particle drag simulations and elucidate how their behaviors differ. We first focus on the YK system in which all particles carry identical charge $q_\text{d}=q_0=1$. In this purely repulsive interacting system, dragging a single particle within the lattice also leads to self-healing of defects---the dragged particle can sneak through the lattice without permanently displacing any other particles~\cite{RN237} (Fig.~\ref{fig:Mechanism_PD}a, Supplementary Movie 6). This elastic response of the colloidal lattice, namely a ``slip mode'', is fundamentally different from other nontrivial defect-healing processes (e.g., ``double-row'' healing phenomenon), in which the lattice undergoes plastic deformation as one or multiple rows of particles will be permanently displaced.

In fact, by increasing the charge of the dragged particle from $q_\text{d}/q_0=1$ to $q_\text{d}/q_0=5$, we can observe a variety of plastic responses of the lattice: double-row healing of defects is found for $\Psi_d\approx 0^\circ$ and single-row healing of defects is found for $\Psi_d\approx 30^\circ$ (Fig.~\ref{fig:Mechanism_PD}b). The effectively enlarged core of the dragged particle enhances its interaction with the surrounding particles, melts the local lattice, and thereby disrupts the slip mode by stabilizing a packet of particles that are transported together~\cite{RN237} (Supplementary Movie 7). This collective motion of particles poses minimal destruction to the lattice by healing all defects in the wake of their trajectory (except the initial and current locations of the packet). 

Despite the similarity of this phenomenon to the AVM model, the microscopic details of the defect-healing process differ between the two models. In the YK model, the co-moving particle(s) is (are) positioned in front of the dragged particle, whereas in the AVM model, the co-moving cell follows the dragged cells. Dragging a particle of the same size ($q_\text{d} = q_0$) in the YK model likely results in the slip mode. In contrast, dragging a single cell in the AVM model does not yield slip mode; because the cells are highly adhesive, this stickiness prevents clean elastic slipping and instead leads to a non-healing, destructive mode.

Next, we analyze the LJ potential. We find that dragging a sticky particle with $\epsilon_\text{d0}/\epsilon_\text{00}\approx7.746$ promotes multi-particle transport (Fig.~\ref{fig:Mechanism_PD}c). 
For large $\Psi_d$($\ge20^\circ$), single-row displacement mode is disrupted, leading to non-healed defects. However, stickiness favors defect healing for small $\Psi_d$($<20^\circ$). A further enhancement of stickiness promotes self-healing of defects via multi-row displacement of cells (Fig.~S20, Supplementary Movie 8). This parallels the multicellular streaming observed in cancer cell clusters, where short-range cell--cell attraction is identified as a key driver of collective migration~\cite{RN101}, suggesting that interparticle attraction promotes directed collective displacements of cells.
Comparing the LJ model to the AVM model, however, we find that defect-healing is suppressed in the AVM model (Fig.~\ref{fig:Mechanism_PD}d). Despite that both models have adhesion or attractive force between particles/cells, the AVM and LJ model are different in three aspects: (1) AVM cells have fixed areas, but LJ particles do not have a well-defined area; (2) AVM cells adhere to each other as long as they share edges, but LJ particles become less attractive as they are separated; (3) Many-body interactions---such as T1 transitions---are important dynamic processes in the AVM model, but are absent in the LJ model. These differences complicate the cell--cell interactions in the AVM model, destabilize the core cells, and lead to less ordered lattice structure in the wake of the dragged cells. Similarly, self-healing of defects is also difficult to observe in the NexTissUe model, in which core cells are even less adhesive than those in the AVM model. In the NexTissUe model, void space behind the dragged cells commonly exists (Supplementary Movie 9), whereas the AVM model enforces confluence and the follower cell strongly adheres to the dragged cells, stabilizing the core structure.

%Introducing long-range interparticle attraction via the LJ potential ($\epsilon_\text{d0}/\epsilon_\text{00}=3$) further facilitates multi-row plastic deformation modes (Fig.~\ref{fig:Mechanism_PD}c): the attraction disrupts the single-row mode near $\Psi_d=30^\circ$ and favors multi-row healing with displaced row number $n \geq 3$ for high $\epsilon_\text{d0}/\epsilon_\text{00}$ (Fig.~S18). Notably, the multi-row self-healing in the LJ system is accompanied by local streaming of surrounding particles near the dragged core (Fig.~S18) --- a directional collective flow emerging from the same short-range attraction. This parallels the multicellular streaming observed in cancer cell clusters, where short-range cell--cell attraction is identified as a key driver of collective migration~\cite{RN101}, suggesting that interparticle attraction promotes directed collective displacement. In confluent tissue, cell--cell attraction is strictly local --- confined to the two immediate neighboring cell lines --- and the LJ-type enhancement of healing is therefore restricted to the double-row configuration. Peculiar to confluent tissue systems, it is the many-body interactions among cells that stabilize the multi-cell core against dissolution and sustain the self-healing packet, giving rise to richer physical behaviors than those accessible in colloidal systems, as discussed in the preceding section (Fig.~\ref{fig:Mechanism_PD}d).

In the defect-healing phenomena observed in all models, the Burgers vectors of the moving packet amount to zero. There are two types of cells or particles in the packet. If $n$ cells or particles are dragged, there can be $m$ dynamic cell(s)/particle(s) with displacement vectors of unit-cell length and $l\ge n$ cell(s)/particle(s) transported with large displacement vectors. For a multi-row defect-healing scenario, the number of rows is exactly $m$. This is because that the collective unit-cell streaming of a row of cells can contribute to one dynamic addition of cell to the moving packet. 
The $l$ cells/particles contribute to the extra cells/particles found in the packet region, within which defects are inevitable. $l=n$ can be observed in the YK model in which particle adhesion is absent. In other models containing particle/cell adhesion forces, $l>n$ holds. This means that there will be $l-n$ cell(s)/particle(s) closely following the dragged ones, and these $l$ cells/particles comprise the core of the moving packet. Both the number $l$ and the core structure vary among different models. The AVM and NexTissUe models favor ``follower cell(s)'', which can stick to the back of the two dragged cells. In the colloidal systems, by contrast, ``pushed particle(s)'' in front of the dragged particle are frequently observed in the LJ model. Occasionally, even two lining-up particles can be observed in front of the dragged one (Fig.~S20, Supplementary Movie 10). These pushed particles do not drop off the packet. These interesting differences between models reveal the subtle competition among cell adhesion forces, many-body interactions, and other factors (e.g., fixed shape versus flexible shape).

%The fixed and dynamic additional cell decomposition introduced for the AVM packet applies universally across lattice systems: the number of self-healing rows determines the number of dynamic additional particles, while the fixed core composition varies with the interaction potential. In the YK and LJ colloidal systems, the flux of the displacement field at the packet boundary can likewise be decomposed into short-range contributions from the dynamic cells and long-range contributions from the fixed core. Beyond the dragged particles, however, the nature of the fixed core differs fundamentally between confluent tissue and colloidal systems. In the AVM and NexTissUe models, many-body interactions favor a ``follower cell'' that attaches behind the doubly-dragged cells. In colloidal systems, by contrast, ``pushed particles'' in front of the dragged particle are frequently observed for the LJ potential in particular, lining up stably and sometimes accumulating more than one (Fig.~S18, Supplementary Movie 6). These pushed particles do not drop back, creating a carrier configuration opposite to that of confluent tissue systems. Consequently, the number of additional cells in the LJ packet is significantly greater than in the YK system, highlighting the significance of short-ranged adhesion in global migration.

Taken together, these results reveal a unified physical picture: purely repulsive systems favor an elastic lattice response, where all particles return to their original positions and hexatic order is fully restored. The introduction of an enlarged dragging core, short-range interparticle attraction, or -- uniquely in confluent tissue -- T1 transition energy barriers disrupts this elastic response and gives rise to nontrivial plastic deformation that permanently displaces cell rows while preserving hexatic order. Notably, the $0^\circ$ direction emerges as a preferred channel for collective migration: it is the unique direction along which nontrivial plastic deformation -- and thus directed row displacement -- is sustained. In a tissue residing in the hexatic phase, where local orientational order extends over many cell lengths, this directional selectivity can couple to the underlying lattice symmetry and channel collective cell motion into directed streams, representing a physically significant anisotropy in an otherwise globally disordered tissue.

\section{\label{sec:discusion}Discussion}

In this work, we have used the AVM model to systematically investigate the dynamic patterns of topological defects in confluent cells. We first examine the statistics of different types of structural defects, and find that a higher activity $\bar{v}_0$ not only liquefies the system but also differentiates $+$ defect, $-$ defect, and hexagonal cells, promoting global fluctuations as well as collective cell migration, and a larger shape parameter $p_0$ also liquefies the confluent cells but homogenizes different types of cells.
We also demonstrate that the age and lifetime of topological defects contain important information about the structure and dynamics of confluent cells. For example, by tracking nematic defects in the system, we find that they exhibit stronger extensile character as they age, as they are more likely to unbind from each other after birth.

By probing the connection between cell-morphology dictated nematic order and cell-topology dictated lattice structures, we establish a spatio-orientational correlation between $+1/2$ defects and structural defects. The two defect types are geometrically coupled through the local cellular arrangement, as confirmed in published experiments as well as in other simulation models~\cite{RN10, RN25, RN4, RN115}. A recent analysis on MDCK monolayers has claimed that nematic ($\pm1/2$) and hexatic ($\pm1/6$) defects measured from the coarse-grained field of the tissues show no significant positional or orientational coupling between them~\cite{RN245}. We believe that this seemingly different result is due to different definitions of defects. in Happel et al., hexatic defects are $\pm1/6$ singularities in the coarse-grained $\psi_6$ orientational field --- field-level objects whose location depends on the coarse-graining scale~\cite{RN245}. However, our $+/-$ structural defects are cell-level objects, namely specific cells whose neighbor count $z$ deviates from 6. Furthermore, their experiments focus on the fluid phase, whereas our analysis targets the solid and hexatic phases in which the system is dynamically arrested. Therefore, cell coordination directly characterized by the structural defects is highly correlated with its deformation, which is captured by the nematic defects in the coarse-grained field.

%A recent experimental study measured correlations between coarse grained nematic ($\pm1/2$) and hexatic ($\pm1/6$) defects in MDCK monolayers, finding no significant positional or orientational coupling between them~\cite{RN245}, which appears to contrast with our finding of a clear spatio-orientational correlation between $\pm1/2$ defects and structural defect cells. The discrepancy, however, stems from a fundamental difference in defect definition: in Happel et al., hexatic defects are $\pm1/6$ singularities in the coarse-grained $\psi_6$ orientational field --- field-level objects whose location depends on the coarse-graining scale --- whereas our structural defects are cell-level objects, namely specific cells whose neighbor count $z$ deviates from six. Furthermore, their experiments focus on the fluid phase, whereas our analysis targets the solid and hexatic phases where the system is not fully melted and the T1 transition energy barrier plays a non-negligible role, rendering cell coordination a geometrically and mechanically meaningful quantity. The two approaches therefore probe different levels and regimes of tissue organization and are complementary rather than contradictory.

By simulating a cell dragged in a hexagonal lattice in the solid state, we observe a dynamic defect pattern surrounding the migrating cell and in the wake of its trajectory. We further identify anisotropic migrations under the 6-fold symmetry where cells experience the least drag at $\Psi_\text{d}=0^\circ$ and the highest drag at $\Psi_\text{d}=30^\circ$. There is a weak directional locking effect as cells tend to migrate at $0^\circ$, at which the dragged cell experiences the least drag by managing to locomote between two rows of cells. 

The single-cell drag simulation can be interpreted as an active microrheology measurement, where the dragged cell functions as a driven probe requiring a critical force $\bar{F}_\text{c}$ to escape from its local cage. This critical force is quantitatively linked to the macroscopic yield stress of the system~\cite{RN179}, and this is consistent with the yielding behavior observed in the macrorheological analysis of solid-state confluent cells with low $\bar{v}_0$ and $p_0$~\cite{RN7}. Beyond $\bar{F}_\text{c}$, the probe fluidizes its surrounding microenvironment, inducing a finite effective viscosity. The post-yielding force--velocity relationship exhibits a linear regime ($\bar{F}_\text{d}>\bar{F}_\text{c}$), which is analogous to the microrheological response of a homogeneous viscoelastic medium~\cite{RN39}. By introducing the apparent viscosity as $\eta_\mathrm{app}\propto\frac{\bar{F}_\text{d}}{\bar{v}}$, the system demonstrates a weak shear-thinning behavior above the critical force (Fig.~\ref{fig:4-1}). 

Recent theory has established that structural defects are intimately connected to collective cell migration~\cite{Krommydas_2025}. Here we demonstrate a complementary result: dragging two neighboring cells can autonomously heal the structural defects in their wake while sustaining persistent migration, preserving the structural integrity of the tissue during their migration. This self-healing is accompanied by a compact neutral defect packet with vanishing Burgers vector. The free energy analysis reveals a nontrivial period of $T_{f_1} = 2.5T_0$, and we identify a ``pedestrian''-like walking pattern: the two winglet cells advance alternately in a ``2+3'' and ``3+2'' manner--- this explains the nontrivial period $5/2$. During this process, a follower cell adheres to the two dragged cells, forming a stable three-cell core. The spatial distribution of T1 transitions (Fig.~S18a) further reveals two sharply localized heat zones. This pattern is absent in the single-cell drag simulation, where scattered T1 events produce unhealed structural defects in the wake. The defect distributions further confirm this trend: the double-drag case shows a compact wake, whereas the single-cell case exhibits a broadened, ``flame-like'' defect pattern, especially at larger drag angles (SI, Fig.~S19).

To understand the physical origin of the self-healing phenomenon and test its generality, we have tested various cell/particle drag simulations using fundamentally different model systems, including the Yukawa colloidal system, the Lennard--Jones colloidal system, and the NexTissUe model. In the purely repulsive YK system, the slip mode dominates. Electrically repulsive particles can slide by each other without adhesion. As a result, the whole lattice remains unchanged while the particle is migrating. However, enlarging the core size of the dragged particle in the YK model facilitates more coordinated particle migration by effectively enhancing stickiness between particles, and defect-healing behaviors similar to those observed in the AVM model can be observed~\cite{RN237}. Introducing a short-range attraction in the LJ model further confirms that particle--particle adhesion promotes larger moving cores and favors multi-row displacements of particles (Fig.~S20). Compared to the colloidal particle models, however, a full defect annealing is harder to observe in the two cell models. This can be attributed to cell confluence and many-body interactions, which can tolerate lattice irregularities by constraining cell area and cell contacts while promoting cell rearrangements via T1 transitions.

%breaks this elastic response, as the enlarged core locally melts the surrounding lattice~\cite{RN237}, disrupting the slip mode and enabling the neutral defect packet that mediates the nontrivial plastic row displacement while holding the hexatic integrity. This universality is confirmed across all model systems studied --- Yukawa, Lennard--Jones, AVM, and NexTissUe --- which differ fundamentally in interaction potentials, cell representations, and T1 transition mechanisms, yet all exhibit the same self-healing phenomenology rooted in hexatic solid mechanics. 

%Beyond the double-row mode, short-range attraction in the LJ system and local adhesion in confluent tissue unlock a richer family of multi-row ($n \geq 3$) self-healing configurations. With increasing attraction strength, the number of collectively displaced rows grows, and local streaming of particles near the dragged core emerges (Fig.~S18%\ref{fig:multi-line}
%), paralleling the multicellular streaming observed in cancer cell migration through dense tissue~\cite{RN101}. Crucially, the $0^\circ$ direction emerges as the uniquely preferred channel for plastic deformation across all systems, coupling the self-healing mechanism directly to the hexatic lattice symmetry and channeling collective migration into directed streams.

Our cell drag simulations have biological relevance.
Collective cell migration, a cornerstone of biological processes such as tissue repair, embryonic morphogenesis, and cancer metastasis~\cite{RN173, RN174, RN175, RN101, RN251}, frequently adopts a \textit{leader-follower} topological configuration to sustain persistent directional movement~\cite{RN168}. While prior studies have predominantly focused on biochemical signaling as the regulatory framework for such collective behavior---particularly in contexts like wound healing and tumor invasion~\cite{RN166}---our findings highlight the mechanical origin. We demonstrate that microscale self-healing \textit{leader-follower} architecture, without any biochemical guidance cues, is sufficient to drive ordered collective migration through purely mechanical interactions. This mechanical perspective complements existing biochemical models, suggesting that the physical arrangement of cells and their force transmission dynamics can alternatively facilitate cell migration even in the absence of biochemical signaling pathways. 

In summary, our work provides new physical insight\textcolor{blue}{s} for understanding the complex dynamics of active confluent tissues and revealing the roles of topological defects in the steady state and during cellular transport. Importantly, we have predicted a possible way of defect-free cell migration without any biochemical coordination between cells. During this cell migration process, defect cells annihilate in a nontrivial spatiotemporal pattern. By comparing defect-healing phenomena between different models, we uncover the roles of cell adhesion, cell confluence, and many-body interactions during defect healing processes.
The predicted defect-healing migration of cells can be heuristic for the further understanding of cellular transport during many physiological processes, such as wound healing and cancer metastasis, in which stem cells and cancer cells can migrate while preserving tissue integrity.
Future works can be devoted to three-dimensional simulations as well as how biochemical factors and extra cellular matrix can contribute to the collective transport of cells.

\section{\label{sec:acknowledgement}Acknowledgement}
R.Z. acknowledges support from the Research Grants Council of Hong Kong via grant number 16300221, C6001-23Y, and NSFC--RGC Joint Research Scheme N\_HKUST627/23. B. L. acknowledges support from National Natural Science Foundation of China via grant no. 12361161604.
We gratefully acknowledge inspiring discussions with Prof. Tiezheng Qian on solid-liquid transitions and valuable inputs from Prof. Greg Huber regarding drag simulation. We are indebted to Prof. Rastko Sknepnek for his expert guidance on lattice statistics and suggestions for model refinement, and to Prof. Dapeng Bi for fruitful discussions on structural defect annihilation. Special thanks go to Dr. Lakshmi Balasubramaniam for providing critical experimental evidence and engaging discussions, and to Prof. Luca Giomi for helpful and encouraging suggestions. We also extend our appreciation to Prof. Yun Chang for sharing valuable biological perspectives on our results, and to Profs. Liting Duan and Qun-Li Lei for their constructive suggestions and insights. We are grateful to Prof. Massimo Pica Ciamarra for valuable discussions and for generously providing the NexTissUe model code. Finally, we thank our group members Zeyang Mou, Wentao Tang, Haijie Ren, and Zeming Liu for their stimulating discussions and contributions throughout this work.

\section{\label{authorcontributions}Author contributions}
R.Z. conceived and supervised the project; J.Z. and C.W.C. developed the computational framework; J.Z. conducted simulations and data analysis; J.Z. and R.Z. wrote the manuscript; B.L. participated discussion and edited the manuscript.
\section{\label{authordeclaration}Author declaration}
No competing interests.

\bibliography{main}

\end{document}